\documentstyle[11pt]{article}

\begin{document}

\begin{center}
\large {\bf State Measurements with Short Laser Pulses \\
and Lower-Efficiency Photon Detectors} \\
\def\thefootnote{\fnsymbol{footnote}}
\vspace*{.5cm}
   Almut Beige\footnote{e-mail: beige@theorie.physik.uni-goettingen.de}
   and Gerhard C. Hegerfeldt\footnote{e-mail: 
   hegerf@theorie.physik.uni-goettingen.de}  \\ [.5cm]
\normalsize
   Institut f\"ur Theoretische Physik \\
   Universit\"at G\"ottingen\\
   Bunsenstr.~9\\
   D-37073 G\"ottingen, Germany
\end{center}
\setcounter{footnote}{0}

\vspace*{0.5cm}

\begin{abstract}
It has been proposed by Cook ({\em Phys.~Scr.}~T {\bf 21}, 49 (1988))
to use a short probe laser pulse for state measurements of two-level 
systems. In previous work we have investigated to what extent this
proposal fulfills the projection postulate if ideal photon detectors
are considered. For detectors with overall efficiency less than 1 
complications arise for single systems, and for this case we present a 
simple criterion for a laser pulse to act as a state measurement and to 
cause an almost complete state reduction.
\end{abstract}
\vspace*{0.5cm}

\section{Introduction}
The outcome of measurements is often described in an idealized way by
the projection postulate with its state vector reduction. The current 
form of the projection postulate is due to L\"uders \cite{Lue} and 
for degenerate eigenvalues it differs from that of von Neumann \cite{Neu}.
For non-degenerate eigenvalues it is implicitly contained in Dirac's
book \cite{Dirac}.
For the simple case of a two-level system in a state $|
\psi\rangle$ and for a measurement of the energy the projection
postulate states that, with
probability $|\langle1 | \psi\rangle|^2$, one finds $E_1$ and the state is
changed
to $|1\rangle$; correspondingly for $E_2$. In this simple case a measurement
of the energy can also be regarded as a measurement of the eigenstates
$|1\rangle$ and $|2\rangle$.

Cook \cite{Cook} proposed to perform a measurement of the states $|1\rangle$
and $|2\rangle$ of such a two-level system by means of a short laser pulse
which drives the transition between level 1 and an additional, rapidly
decaying, level 3 (see Fig.~1). For a single system, initially in a
superposition of $|1\rangle$ and $|2\rangle$, the short laser pulse may, or may
not, lead to the emission of fluorescence photons. In the former case
the system was taken to be in $|1\rangle$ after the end of the pulse (and
after a short transient decay time for level 3), since right after
each photon emission it is in the ground state. In the latter case the
system was taken to be in $|2\rangle$. This was used in an experiment
on the quantum Zeno effect \cite{Wine}.

In two previous papers \cite{BeHe,BeHeSo} we have investigated in how
far this proposal realizes a measurement to which the projection
postulate can be applied. Related work appeared in 
Refs.~\cite{Petr,Block,Fre,Gag,Knight,Mahler}. 
It was shown in Ref.~\cite{BeHe}
that, under the condition of no photon emission during a pulse, the reduced (or
conditional) time development \cite{HeWi,Wi,He,HeSo,QT} carries the state of
the system ever closer to $|2\rangle$. The quality of this reduction to 
$|2\rangle$ can be expressed by a single combination of the pulse
parameters, namely $N_{\rm e}(\tau_{\rm p}; |1\rangle)$, the average number of
photons emitted
during a pulse of length $\tau_{\rm p}$ when one starts from $|1\rangle$ as
initial state. $N_{\rm e}$ can be calculated from the fluorescence rate of
an ensemble of such systems. It was shown in Ref.~\cite{BeHe} that for
$N_{\rm e}$ as low as 5 or 10 one has an excellent reduction to $|2\rangle$. 

In quantum mechanics it is in principle not legitimate to speak about
emission of a photon unless it is observed (detected). However, it was
pointed out in Ref.~\cite{HeWi} and Ref.~\cite{Reib} that in questions
of photon statistics it makes no difference theoretically whether or
not the photons are monitored over all space at rapidly repeated time
intervals. Emitted photons are then those which would be
detected by ideal detectors in the whole space. 

The discussion in Refs.~\cite{BeHe,BeHeSo} is based on such ideal
detectors. With non-ideal detectors problems arise in the reduction of
the state of a single system when no fluorescence photons are detected
during a pulse, because this null detection may be due either to no
emission at all or to a failure of the detector to record emitted
photons. In the former case the system is in $|2\rangle$ at the end of the
pulse, while in the latter case it is in $|1\rangle$. Hence the system has
to be described by a density matrix with a $|2\rangle\langle2|$ and a 
$|1\rangle\langle1|$
contribution. For a good reduction to $|2\rangle$ the latter, the
``non-reduced part'' should be small, and the aim of this paper is to
give a simple criterion to ensure this. 

A real detector may miss a photon because of low quantum efficiency or
because it operates only in a smaller solid angle\footnote{
Solid angles smaller than $4\pi$ have been considered in
Ref.~\cite{PleDiss,HePle6}, where in one solid angle photons were
counted, while in the other solid angle the light was spectrally
analyzed.}. We define the overall efficiency $\eta'$ as the ratio of
detected vs. emitted photons. 

For a system initially in the ground state, we denote by 
$N_{\rm d}(\tau_{\rm p};|1\rangle) \equiv \eta' N_{\rm e}(\tau_{\rm p}; 
|1\rangle)$ the average number of
fluorescence photons detected during a pulse. Experimentally this is
easy to determine from the observed fluorescence of an ensemble
(``gas'') of systems, and one can adjust the pulse parameters to
realize any value of $N_{\rm d}$ for a given detector. In a
simplified derivation we will show in Section 3 of this
paper that for initial state $|\psi\rangle$ the non-reduced part is 
approximately
\begin{equation}\label{I1}
|\langle1|\psi\rangle|^2 ~ {\rm e}^{-N_{\rm d}}
\end{equation}
for small overall efficiency $\eta'$. However, this simple derivation
gives no insight on how good the approximation is and how 
small $\eta'$ has to be. Therefore we
show in the precise treatment of Section 4 that the 
non-reduced part is smaller than
\begin{equation}\label{I2}
|\langle1|\psi\rangle|^2 (1 +~\frac{2}{3}~\eta')~ {\rm e}^{-N_{\rm d}}
\end{equation}
for $\eta' \leq 1/3$ and $N_{\rm d} \geq 1$, provided $N_{\rm e}$ is 
sufficiently large, e.g. 5, to ensure reduction in
the case of ideal detectors. This is a strict bound from above;
a strict bound from below is given at the end of Section 4 in Eq.
(\ref{H44}). If the initial state is a density matrix
$\rho$ instead of a pure state $|\psi\rangle$, then 
$|\langle1|\psi\rangle|^2$ is replaced by $\rho_{11}$. We derive 
Eq.~(\ref{I2}) by Laplace transforms. Similar techniques have been used 
in Ref.~\cite{Car}. 

\section{Basics and previous results}
We consider a V system as in Fig.~1, with level 2 (meta-)stable and
level 3 rapidly decaying (Einstein coefficient $A$). The laser pulse has
Rabi frequency $\Omega$ and is tuned to the 1--3 transition. In the
quantum jump approach (or Monte-Carlo wave function approach, or
quantum trajectories) \cite{HeWi,Wi,He,HeSo,QT,MC} it is shown that
under the condition of no photon emission the time development of a
system is given by a reduced (or conditional) Hamiltonian. In our case
it is given in an interaction picture by \cite{BeHe}
\begin{equation}\label{2}
H_{{\rm red}} \equiv -{\rm i} \hbar M =~\frac{\hbar}{2}~\left\{ \Omega (
|1\rangle\langle3| + |3\rangle\langle1|) - {\rm i} A~|3\rangle\langle3|
\right\} ~.
\end{equation}
The eigenvalue of $M$ are $\lambda_2 = 0$ (with eigenvector $|2\rangle$) and
\begin{equation}\label{3}
\lambda_{1,3} = ~\frac{1}{4}~ \left( A \mp~\sqrt{A - 4 \Omega^2} \right)~.
\end{equation}
For initial state $|\psi\rangle$,
\begin{equation}\label{H9}
|\psi\rangle = \alpha_1 |1\rangle + \alpha_2 |2\rangle~,
\end{equation}
the conditional time development is then 
\begin{equation}\label{6}
{\rm e}^{-i H_{{\rm red}} \tau/\hbar} |\psi\rangle 
= \alpha_2 |2\rangle + \alpha_1 ~\left\{ 
\frac{\lambda_1 {\rm e}^{- \lambda_3 \tau} -\lambda_3 {\rm e}^{-\lambda_1
\tau}}
{\lambda_1 - \lambda_3} |1 \rangle 
+\frac{{\rm i}}{2}~\Omega \frac{{\rm e}^{-\lambda_1 \tau}
- {\rm e}^{-\lambda_3 
\tau}}{\lambda_1 - \lambda_3} |3\rangle\right\}~.
\end{equation}
The probability for no photon emission until time $\tau$, denoted by
$P_0^{\rm e} (\tau; |\psi \rangle )$, is given by
\begin{equation}\label{5}
P_0^{\rm e} (\tau; |\psi \rangle ) 
= \| {\rm e}^{- {\rm i} H_{{\rm red}} \tau/\hbar}
|\psi \rangle \|^2~.
\end{equation}
 From Eq.~(\ref{5}) one obtains
\begin{equation}\label{6a}
P_0^{\rm e}(\tau; |\psi \rangle) = |\langle 2 | \psi \rangle |^2 + | \langle
1 |\psi \rangle |^2 P_0^{\rm e} (\tau; |1 \rangle )~.
\end{equation}
The probability density for the first photon emission is 
\begin{eqnarray}\label{10a}
w^{\rm e}_1(\tau; |\psi \rangle ) & = & -~\frac{{\rm d}}{{\rm d} \tau}~ 
P_0^{\rm e} (\tau; | \psi\rangle )\nonumber\\
& = & | \langle 1 |\psi \rangle|^2 w^{\rm e}_1 (\tau; |1 \rangle)
\end{eqnarray}
since the $|\langle 2|\psi\rangle|^2$ term in Eq.~(\ref{6a}) drops out.
In particular, for an atom with initial state $|1 \rangle$ one has
\begin{equation}\label{9}
P_0^{\rm e} (\tau;|1\rangle) = ~\frac{1}{4}~\Omega^2
\left( \frac{{\rm e}^{-\lambda_1 
\tau} - {\rm e}^{-\lambda_3 \tau}}{\lambda_1 - \lambda_3} \right)^2 + 
\left( \frac{\lambda_1 {\rm e}^{-\lambda_3 \tau} - \lambda_3
{\rm e}^{-\lambda_1 
\tau}}{\lambda_1 - \lambda_3} \right)^2
\end{equation}
and
\begin{equation}\label{10}
w_1^{\rm e}(\tau;|1\rangle) = ~\frac{1}{4}~\Omega^2 A \left(
\frac{{\rm e}^{-\lambda_1 \tau} - {\rm e}^{-\lambda_3 \tau}}{\lambda_1 -
\lambda_3} \right)^2~.
\end{equation}

If $\tau_{\rm p}$, the duration of the pulse, satisfies 
\begin{equation}\label{7}
\tau_{\rm p} \gg \max \{ 1/A~;~A/\Omega^2 \}
\end{equation}
then for $\tau = \tau_{\rm p}$ the exponentials in Eq.~(\ref{6}) can be
neglected and the state is proportional to $|2 \rangle$, with
probability
\begin{equation}\label{5a}
P^{\rm e}_0 (\tau_{\rm p}; |\psi \rangle) = | \langle 2 | \psi \rangle |^2~,
\end{equation}
just as with the projection postulate. In this way one can understand
why a system without emissions during a pulse ends up in the state $|2
\rangle$.

With the complementary probability $1 - |\langle 2 | \psi \rangle |^2
= |\langle 1 | \psi \rangle |^2$ an atom will emit photons during a
pulse and will be in a superposition of $| 1 \rangle$ and $| 3
\rangle$ at its end, at time $\tau_{\rm p}$. The $| 3 \rangle$ component
will then either lead to the emission of an additional photon or the
reduced time development (now with $\Omega = 0$) leads to its
vanishing (without emission). In both cases the state will be in
$| 1 \rangle$, after a transient time $\tau_{{\rm tr}}$ whose duration
is several multiples of $A^{-1}$. 

Analogously, 
for a system with no emission until $\tau_{\rm p} + \tau_{{\rm tr}}$
the tiny $| 3 \rangle$ component, still present at time $\tau_{\rm
p}$, has vanished by then and the
remaining  tiny $| 1 \rangle$ component of $\exp \{-{\rm i} H_{\rm red}
\tau_{\rm p}/\hbar \} | \psi \rangle$ in Eq.~(\ref{6}) is the part which 
is not reduced to $| 2 \rangle$. This can be estimated more precisely in 
terms of $N_{\rm e} (\tau_{\rm p} ; | 1 \rangle)$. The latter is obtained by
integrating
the photon rate $A \rho_{33}(\tau)$ where $\rho_{33}(\tau)$ is given
by the Bloch equations for the driven 1--3 two-level system with
$\rho_{11}(0) = 1$ \cite{London}. This gives 
\begin{equation}\label{18}
N_{\rm e} (\tau;|1\rangle) =~\frac{\Omega^2}{A^2 + 2\Omega^2}~\left(
A\tau~-~\frac{3 A^2}{A^2 + 2 \Omega^2}~+~\frac{2 A}{A^2 + 2
  \Omega^2}\cdot\frac{\mu^2_1 {\rm e}^{-\mu_2 \tau} - \mu^2_2 {\rm e}^{-\mu_1
      \tau}}{\mu_1 - \mu_2}~\right)
\end{equation}
where
\begin{equation}\label{19}
\mu_{1/2} =~\frac{1}{4} \left( 3A \mp \sqrt{A^2 - 16\Omega^2} \right)~.
\end{equation}
For $N_{\rm e} \ge 1$ the contribution of the exponential terms is 
negligible. (For weak driving the time for the first emission is long
which makes $\exp (-\mu_i \tau)$ small; for strong driving the factor in
front of the last term is small.) We can and will therefore use, for 
$N_{\rm e} \ge 1$,
\begin{equation}\label{18a}
N_{\rm e} (\tau_{\rm p};|1\rangle) =~\frac{\Omega^2}{A^2 +
2\Omega^2}~\left( A\tau_{\rm p} -~\frac{3 A^2}{A^2 +2\Omega^2}\right)~.
\end{equation}
With this one can express $\tau_{\rm p}$ by $N_{\rm e}(\tau_{\rm 
p};|1\rangle)$ (and the remaining parameters) and can insert it for 
$\tau$ on the r.h.s. of Eq.~(\ref{6}). A graphical evaluation then
yields as upper bound for the non-reduced part
\begin{equation}\label{21}
|\langle 1 | \exp \{- {\rm i} H_{{\rm red}} \tau_{\rm p}/\hbar\} 
| \psi \rangle|^2~\leq |\langle 1|\psi \rangle|^2 ~ 
{\rm e}^{-N_{\rm e}(\tau_{\rm p};|1\rangle )}~.
\end{equation}
Thus in the case of no photon emission during a pulse the reduction of
$| \psi \rangle$ to $| 2 \rangle$ is already excellent if the
parameters of the pulse are such that $N_{\rm e}(\tau_{\rm p};| 1 \rangle)
\geq 5$,
i.e.~if the average number of photons emitted during the pulse by an
atom initially in the ground state is 5 or more. The estimate in
Eq.~(\ref{21}) can be improved substantially for strong driving
\cite{BeHe}.

\section{Simplified derivation}
We now incorporate a non-ideal detector of overall efficiency
$\eta'<1$. With initial state $|\psi
\rangle$ as in Eq.~(\ref{H9}) we denote by $P^{\rm d}_0(\tau;| \psi
\rangle)$ the probability of no {\em detection} of fluorescence
photons until time $\tau$ for a single system. With probability
$P^{\rm e}_0(\tau_{\rm p}; | \psi \rangle)$ no photon is emitted during
a pulse,
and we assume in the following that for ideal detectors the non-reduced
part is negligible, i.e. the corresponding state is $| 2
\rangle$. With probability $P^{\rm d}_0 - P^{\rm e}_0$ photons are emitted, but
missed by the non-ideal detector, and in this case the state is $| 1
\rangle$, after a transient decay time $\tau_{\rm tr}$ for the 3
component. Hence a system, for which no fluorescence photons are
detected, is described by the density matrix
\begin{equation}\label{H1}
\rho_{\rm d}^0(\tau_{\rm p};\eta') \equiv 
P^{\rm e}_0 (\tau_{\rm p} ; | \psi \rangle) ~| 2 \rangle \langle 2 | + 
\left( P^{\rm d}_0
(\tau_{\rm p} ; | \psi \rangle) - P^{\rm e}_0 (\tau_{\rm p} ; |\psi 
\rangle) \right) ~| 1 \rangle \langle 1 |
\end{equation}
where the normalization is such that the trace is $P^{\rm d}_0$, i.e. the
probability of no detection. We call the $|1\rangle\langle 1|$ 
component the non-reduced
part, and the smaller this is the better a reduction as in the
projection postulate is realized. If one makes no photon
measurements, i.e. $\eta'=0$ and $P_{\rm d} \equiv 1$, then 
$\rho_{\rm d}^0(\tau_{\rm p};\eta') $ is just the diagonal density
matrix of the Bloch equations which was discussed in Refs.
\cite{Block,Fre}.  

We will show further below (Section~4.1) that for $P^{\rm d}_0$ an equation
analogous to Eq.~(\ref{6a}) holds, 
\begin{equation}\label{H4}
P^{\rm d}_0 (\tau;|\psi\rangle) = |\langle 2 | \psi \rangle |^2 + | \langle
1 | \psi \rangle |^2 P^{\rm d}_0 (\tau;|1\rangle)~.
\end{equation}
The density matrix in (\ref{H1}) then becomes, with Eq.~(\ref{5a}) for
$P^{\rm e}_0$, 
\begin{equation}\label{H1a}
\rho_{\rm d}^0(\tau_{\rm p};\eta') =
| \langle 2 | \psi \rangle |^2 ~|2 \rangle\langle 2| + |\langle 1| \psi
\rangle |^2 P^{\rm d}_0 (\tau_{\rm p} ; | 1 \rangle) ~| 1 \rangle 
\langle 1 |~.
\end{equation}

To determine the non-reduced part one therefore only has to calculate
$P^{\rm d}_0(\tau_{\rm p};|1 \rangle)$, the probability to detect no photon for
initial state $| 1 \rangle$. This will be done in detail in Section
4. Here we use a simplifying assumption, namely that for initial state
$| 1 \rangle$ the fluorescence photons are emitted with a constant
probability rate $I$ during the pulse,
\begin{equation}\label{H5}
I = N_{\rm e} (\tau_{\rm p};|1 \rangle)/\tau_{\rm p}~.
\end{equation}
For the detector the recorded rate is then
\begin{equation}\label{H6}
I_{\eta'} = \eta' I = N_{\rm d}(\tau_{\rm p};|1 \rangle)/\tau_{\rm p}~.
\end{equation}
Similar to radioactive decay Eq.~(\ref{H6}) now implies
\begin{eqnarray}\label{H7}
P^{\rm d}_0 (\tau_{\rm p};|1 \rangle) & = & {\rm e}^{-I_{\eta'}
\tau_{\rm p}}\nonumber\\
& = & {\rm e}^{-N_{\rm d} (\tau_{\rm p}; |1 \rangle)} ~.
\end{eqnarray}
Instead of a constant rate one can also assume the emission process to
be Markovian. This assumption leads to the same result (cf., e.g.,
Ref.~\cite{London}, p.~231).

We expect this to be a good approximation for small $\eta'$ due to
the following. For small $\eta'$ most emitted photons go undetected, and
to have a
given average number $N_{\rm d}$, $N_{\rm d} =5$ say, one needs a large
$N_{\rm e}$. The detected number of photons in
this time interval has then an approximately Poissonian 
distribution, in accordance with Eq. (\ref{H7}). 
However, this argument gives no insight  how small $\eta'$ has to be
or how large the error is. To answer
this we need the more detailed analysis of Section 4.

 From Eqs.~(\ref{H7}) and (\ref{H1a}) one  obtains Eq.~(\ref{I1}).
Thus, with the above simplifying assumption of constant detection rate 
the non-reduced part decreases exponentially with $N_{\rm d} (\tau_{\rm p};
| 1 \rangle)$. For a given detector one can adjust the laser parameters 
accordingly to come as close to the projection postulate as desired. 

\section{General derivation}
\subsection{Solution by Laplace transforms}
With the results of Sections 2 and 3 the incorporation of detectors
with overall efficiency $\eta' < 1$ is a purely statistical
problem. After each emission the system (atom) is in its ground state,
with all memory lost, and therefore the sequence of photon emissions
is a renewal process \cite{Cox}, as is the sequence of photon
detections. This renewal process (although 
derived from the Markovian quantum
trajectories of state vectors) is not Markovian since it would then
be a Poisson process \cite{Cox,Breiman}.
Hence the simplified treatment of Section 3 is only an approximation.

In view of Eq.~(\ref{H1a}) we have to determine $P^{\rm d}_0(\tau_{\rm p} ;
| 1 \rangle)$ in order to estimate the quality of the reduction. For a
system initially in $| \psi \rangle$ the probability to have exactly
$N$ emissions occurring in $(\tau_1, \tau_1 + {\rm d} \tau_1),..., 
(\tau_N, \tau_N + {\rm d} \tau_N)$ in the interval 
$(0,\tau)$ is
\begin{equation}
w^{\rm e}_1 (\tau_1 ; | \psi \rangle) {\rm d} \tau_1
w^{\rm e}_1 (\tau_2-\tau_1 ; | 1\rangle) {\rm d} \tau_2 \cdots 
w^{\rm e}_1 (\tau_N - \tau_{N-1} ; | 1\rangle) {\rm d}
\tau_N P^{\rm e}_0 ( \tau - \tau_N ; | 1 \rangle)
\end{equation}
and hence the probability to have exactly $N$ emissions in $(0, \tau)$,
denoted by $P^{\rm e}_N (\tau;|\psi \rangle)$, is for $N \geq 1$ obtained by
integrating over the domain $0 \leq \tau_1 \leq \tau_2 \leq ...\leq
\tau_N \leq \tau$. With Eq.~(\ref{10a}) for $w^{\rm e}_1 (\tau_1;|\psi
\rangle)$ this gives
\begin{eqnarray}\label{11a}
P^{\rm e}_N (\tau ; |\psi \rangle) &=& |\langle 1 |\psi \rangle
|^2~ \int^\tau_0 {\rm d} \tau_N P^{\rm e}_0 (\tau - \tau_N ;
| 1 \rangle)  ~\int^{\tau_N}_0 {\rm d} \tau_{N-1}
w^{\rm e}_1 (\tau_N - \tau_{N - 1} ; | 1 \rangle) 
\nonumber \\
& &....~\int^{\tau_2}_0 {\rm d} \tau_1
w^{\rm e}_1 (\tau_2-\tau_1; |1\rangle)
w^{\rm e}_1 (\tau_1 ; | 1\rangle)~.
\end{eqnarray}
For $N$ emitted photons the probability for all photons to go undetected 
is $(1- \eta')^N$, and therefore the probability of detecting no photon at
all in $(0, \tau)$ for initial state $| \psi \rangle$ is 
\begin{equation}\label{26}
P^{\rm d}_0 (\tau;|\psi \rangle) = P^{\rm e}_0 (\tau;|\psi \rangle)
+~\sum^\infty_{N=1}~(1 - \eta')^N P_N^{\rm e} (\tau;| \psi \rangle)~.
\end{equation} 
Eq.~(\ref{H4}) immediately follows from this when using
Eqs.~(\ref{11a}) and (\ref{5a}). 

We now use Laplace transform, denoted by $~\hat{ }~$. The convolution 
theorem for Laplace transforms together with Eqs.~(\ref{26}) 
and (\ref{11a}) for $|\psi \rangle =| 1 \rangle$ gives
\begin{eqnarray}\label{27a}
\hat{P}^{\rm d}_0 (p;|1\rangle) & = & \hat{P}^{\rm e}_0 (p;|1\rangle)~
\sum^\infty_{N=0}(1-\eta')^N \hat{w}_1^{\rm e} (p;|1 \rangle)\nonumber\\
&=& \frac{\hat{P}^{\rm e}_0 (p;|1\rangle)}{1 - (1-\eta')
\hat{w}_1^{\rm e}(p;|1\rangle)} ~.
\end{eqnarray}
$\hat{P}^{\rm e}_0$ and $\hat{w}^{\rm e}_1$ are easily obtained from
Eqs.~(\ref{9}) and
(\ref{10}) and yield $\hat{P}^{\rm d}_0$ as a quotient of the form
\begin{equation}\label{27b}
\hat{P}^{\rm d}_0 (p; |1\rangle) =~\frac{g(p)}{f_{\eta'} (p)}
\end{equation}
where
\begin{eqnarray}\label{28}
f_{\eta'}(p) & = & p^3 + ~\frac{3}{2}~A p^2 +~\frac{1}{2}~(A^2 + 2 \Omega^2)
p +~\frac{1}{2} \eta' A \Omega^2 \\ \label{28a}
g (p) & = & p^2 + ~\frac{3}{2} A p +~\frac{1}{2} (A^2 + 2 \Omega^2) =
f_0(p)/p~.
\end{eqnarray}
The same expression was derived by somewhat different methods in
Ref.~\cite{Car} and the zeros of the denominator were found for the
special case $\Omega = A/2$. 

Let $p_1,~p_2,~p_3$ be the zeros of $f_{\eta'} (p)$. Either all of them
are real and negative or one is negative and the other two are complex
conjugates of each other, with negative real parts. We choose $p_1$ to
be the real zero closest to zero and $|\mathop{\rm Re}p_2| \leq 
|\mathop{\rm Re} p_3 |$. By partial fractions one then obtains, for 
distinct zeros,
\begin{eqnarray}\label{30a}
P^{\rm d}_0 (\tau;|1 \rangle)
&=&\sum_{i=1}^3 \frac{g(p_i)}{(p_i-p_{i+1})(p_i-p_{i+2})} ~{\rm e}^{p_i
\tau} \nonumber \\
&\equiv & c_1{\rm e}^{p_1 \tau} + c_2 {\rm e}^{p_2 \tau} + c_3 {\rm
e}^{p_3 \tau}~. \end{eqnarray}
where we use $p_{i+3} \equiv p_i$. If two zeros coincide one has to take
limits.
An alternative form is obtained by noting that
\begin{equation}
f_{\eta'}(p)/(p-p_i) =(p-p_{i+1})(p-p_{i+2})~.
\end{equation}
Letting $p \to p_i$, the l.h.s. becomes $f'_{\eta'}(p_i)=f'_0(p_i)$, and
thus
\begin{eqnarray} \label{30b}
P_0^{\rm d}(\tau;|1\rangle )&=& \sum_{i=1}^3 \frac{g(p_i)}{f'_0(p_i)}~
{\rm e}^{p_i\tau}~.
\end{eqnarray}
Furthermore, since
\begin{equation}
f_0(p)=f_{\eta'}(p)-\frac{1}{2} \eta' A\Omega^2
\end{equation}
and $f_{\eta'}(p_i)=0$, one has from Eq.~(\ref{28a})
\begin{eqnarray} \label{H32}
g(p_i)&=&-\frac{1}{2} \eta' A\Omega^2/p_i~.
\end{eqnarray}
Again using $f_{\eta'}(p_i)=0$ to recast $p_i f'_0(p_i)$ one than finds
\begin{equation}\label{32a}
c_i = ~\frac{g(p_i)}{f'_0(p_i)} ~=~~
\frac{\frac{1}{2} \eta' A \Omega^2}{\frac{1}{2} \eta' A \Omega^2      
- ~\frac{3}{2} A p^2_i - 2 p_i^3}~.
\end{equation}

We will next derive an approximate solution to Eq.~(\ref{30b}) and
will then discuss its range of validity.

\subsection{Approximate form of $P^{\rm d}_0$}
With $\mu_{1,2}$ as in Eq.~(\ref{19}) the zeros of $f_{\eta' = 0} (p)$
are
\begin{equation}\label{16a}
(p_1, p_2, p_3 ) = (0, -\mu_1, -\mu_2)~.
\end{equation}
For small $\eta',~ p_i$ can be found perturbatively by linearization
(Newton's method). In particular one finds
\begin{equation}\label{31b}
p_1 \cong - \eta'~\frac{A \Omega^2}{A^2 + 2 \Omega^2}~.
\end{equation}
This can now be inserted into Eq.~(\ref{32a}) for $c_1$, where in the
denominator we neglect $\eta'^2$ when compared to 1 and $\eta'$.
Eq.~(\ref{18a}) for $N_{\rm e} (\tau_{\rm  p}; | 1 \rangle)$ is used to 
recast  ${\rm e}^{p_1 \tau_{\rm p}}$. With $N_{\rm d} = \eta' N_{\rm e}$
we then obtain
\begin{equation}\label{32b}
c_1 {\rm e}^{p_1 \tau_{\rm p}} =~\frac{1}{1 - 3 \eta' ~\frac{A^2
    \Omega^2}{(A^2 + 2 \Omega^2)^2}}~\exp \left\{ -N_{\rm d} -
3 \eta' ~\frac{A^2
    \Omega^2}{(A^2 + 2 \Omega^2)^2} \right\} ~.
\end{equation}
Expanding in $\eta'$, the linear terms cancel and only $\eta'^2$ terms
with coefficients less than 1 remain; these are neglected when
compared to 1.

 From Eqs.~(\ref{19}) and (\ref{16a}) one sees that, for small $\eta'$,
${\rm e}^{p_2 \tau_{\rm p}}$ and ${\rm e}^{p_3 \tau_{\rm p}}$are much 
smaller in absolute value than
${\rm e}^{p_1 \tau_{\rm p}}$. Therefore the second and third summands in
Eq.~(\ref{30a}) are neglected. (Some care is needed for $p_2$ close to
$p_3$; this is discussed in detail in the next subsection.)

Thus we arrive at the approximate expression for small $\eta'$,
\begin{equation}\label{32c}
P^{\rm d}_0 (\tau_{\rm p} ; |1 \rangle) = {\rm e}^{-N_{\rm d} (\tau_{\rm p}; 
| 1 \rangle)}~.
\end{equation}
In view of the derivation $\eta'$ has to be small enough for
Newton's method to give a reliable $p_1$. In the next subsection the
range of validity of the above expression will be discussed in
detail. The reader not interested in all the details can proceed
directly to the strict bound in Eq. (\ref{H42b}).

\subsection{Validity range and precise estimates}
We now present upper and lower bounds on $P^{\rm d}_0$, and thus on the
non-reduced part. These will justify the preceding approximate solution
in more detail.

For this we will need more information on the zeros of the curve
$y=f_{\eta'} (p)$. This curve is obtained by moving the
curve $y = f_0 (p)$ upward by the amount $\frac{1}{2}~\eta' A
\Omega^2$ or, alternatively, by moving the abscissa downward (see
Fig.~2). We note further that, for $\eta' = 1$, the zeros are,
in the given order,
\[
( p_1, p_2, p_3 ) = \left\{ \begin{array}{lll}
( - 2 \lambda_1, -\frac{1}{2}~A, -2 \lambda_3 )~~~{\rm for}~
\lambda_1~{\rm real}\\
( -\frac{1}{2}~A, -2 \lambda_1, -2 \lambda_3 )~~~{\rm otherwise}
\end{array} \right.
\]
where $\lambda_i$ is given by Eq.~(\ref{3}). The point of inflexion of
the curve is at $p_{\rm infl} = -~\frac{1}{2} A$, which is also a zero of $f_1
(p)$, and this fixes the abscissa for $\eta' = 1$. Thus, from Fig.~2,
$p_1$ lies between $p_{\rm infl}$ and $0$. On the other hand, $p_1$ lies to the
left of the tangent at $p = 0$ (see Fig.~2), and so the r.h.s. of
Eq.~(\ref{31b}), i.e. Newton's method, provides an upper bound for
$p_1$. One also verifies that 
\begin{equation}\label{31c}
- \eta'~\frac{A \Omega^2}{A^2 + \Omega^2}~\leq p_1 \leq -
\eta'~\frac{A \Omega^2}{A^2 + 2 \Omega^2}~~~~{\rm for}~~ 0 \leq \eta'
\leq~\frac{1}{3} \end{equation}
by checking that $f_{\eta'} (p)$ changes sign between these values. 

As zeros of the polynomial $f_{\eta'}$ the $p_i$'s satisfy
\begin{eqnarray}\label{29a}
p_1+p_2+p_3 &=& -\frac{3}{2} A \\ \label{29b}
p_1p_2+p_2p_3+p_1p_3 &=& \frac{1}{2} (A^2 +2\Omega^2)~.
\end{eqnarray}
If $p_2$ and $p_3$ are real they must lie to the left of $p_{\rm infl}$,
i.e. $p_{2,3} \le -~\frac{1}{2}~A$. If $p_1$ is the only real zero
then $p_3 = p_2^*$. Writing 
\begin{equation}\label{29b2}
p_{2/3} = - a \mp {\rm i} b~,~~~a, b > 0
\end{equation}
one obtains from Eqs.~(\ref{29a}) and (\ref{29b})
\begin{eqnarray}\label{H39}
a &=& \frac{3}{4}~A +~\frac{1}{2}~p_1 \\ \label{H40}
b^2 &=& \Omega^2 -~\frac{1}{16}~A^2 +~\frac{3}{4}~p_1^2
+~\frac{3}{4}~Ap_1 ~.
\end{eqnarray}

Now we discuss $c_1 {\rm e}^{p_1 \tau_{\rm p}}$, with $c_1 \ge 0$ given
by Eq.~(\ref{32a}) for $i = 1$. We replace $p_1$ in the exponent and
$p_1^3$ in the denominator of $c_1$ by the r.h.s. of the inequality (\ref{31c})
 and $p_1^2$ by the l.h.s. of (\ref{31c}). Then one obtains instead of
 Eq.~(\ref{32b}), the inequality
\begin{equation}\label{32d}
c_1 {\rm e}^{p_1 \tau_{\rm p}} \le \left\{ 1 - 3 \eta'~\frac{A^2
\Omega^2}{(A^2
+ \Omega^2)^2}~+ 4 \eta'^2~\frac{A^2 \Omega^4}{(A^2 + 2 \Omega^2)^3}
\right\}^{-1} \exp \left\{ - N^{\rm d}_0 - 3 \eta'~\frac{A^2
  \Omega^2}{(A^2 + 2\Omega^2)^2} \right\}
\end{equation}
for $\eta' \le ~\frac{1}{3}$. Evaluating this graphically one finds,
for arbitrary $A$ and $\Omega$ and for $\eta' \le 1/3$
\begin{equation}\label{32e}
c_1 {\rm e}^{p_1 \tau_{\rm p}} \le~\frac{1}{1 - .45 \eta'}~\exp \left\{ 
- N^{\rm d}_0 \right\}~. \end{equation}

For $p_2$ approaching $p_3$ one sees from Eq.~(\ref{30a}) that $c_2$
and $c_3$ diverge, but cancellations occur. To take these into account
we use Eq.~(\ref{H32}) to write
\begin{equation}\label{H3?}
c_2 {\rm e}^{p_2 \tau_{\rm p}} + c_3 {\rm e}^{p_3 \tau_{\rm p}} =
-\frac{\frac{1}{2}~\eta'A \Omega^2}{p_2 p_3 (p_2-p_1)
(p_3-p_1)}~\frac{\{ p_3(p_3 - p_1){\rm e}^{p_2 \tau_{\rm p}} - p_2(p_2-p_1){\rm
e}^{p_3 \tau_{\rm p}}\}} {p_2 - p_3}
\end{equation}
and by recasting the expression in curly bracket as
\begin{equation}\label{H37}
\left\{ (p_3 - p_2)(p_3 + p_2 - p_1) {\rm e}^{p_2 \tau_{\rm p}} + (p_2^2 - p_2
  p_1)({\rm e}^{p_2 \tau_{\rm p}} - {\rm e}^{p_3 \tau_{\rm p}}) \right\}
\end{equation}
we obtain, with Eq.~(\ref{29a}), 
\begin{equation}\label{H38}
c_2 {\rm e}^{p_2 \tau_{\rm p}} + c_3 {\rm e}^{p_3 \tau_{\rm p}} =
-\frac{\frac{1}{2}
\eta'A \Omega^2}{p_2 p_3 (p_2 - p_1)(p_3 - p_1)}
\left\{ \left( \frac{3}{2}A + 2 p_1 \right) {\rm e}^{p_2 \tau_{\rm p}} +
(p^2_2- p_2 p_1)~\frac{{\rm e}^{p_2 \tau_{\rm p}} -
{\rm e}^{p_3 \tau_{\rm p}}}{p_2 - p_3}
\right\} 
\end{equation}
which remains manifestly finite for $p_2 \to p_3$.

If all $p_i$ are real, then it is evident that the expression in the
curly brackets as well as the fraction in front are positive so that
the overall expression is negative. Hence its omission leads in this
case to an upper bound for $P^{\rm d}_0$.

If $p_{2,3}$ are complex, $p_{2,3} = - a \mp {\rm i} b$, then the last
equation can be written as 
\begin{eqnarray}\label{H41}
c_2 {\rm e}^{p_2 \tau_{\rm p}} + c_3 {\rm e}^{p_3 \tau_{\rm p}} 
&=&-\frac{\frac{1}{2}~\eta'A\Omega^2}{(a^2+b^2)(a^2+b^2+2ap_1+p_1^2)}
\nonumber \\
& &\left\{ \left(\frac{3}{2}A+2p_1 \right) \cos b\tau_{\rm p}
- b\sin b\tau_{\rm p} + 
(a^2 + a p_1)~\frac{\sin b\tau_{\rm p}}{b} 
\right\} {\rm e}^{- a \tau_{\rm p}}
\end{eqnarray}
 From Eqs.~(\ref{H39}) and (\ref{18a}) for $a$ and $N_{\rm e}$ one obtains 
\begin{equation}\label{H42}
{\rm e}^{-a \tau_{\rm p}} \le \exp \left\{ - N_{\rm e} - \left(
\frac{3}{4}A+~\frac{1}{2}~p_1 -~\frac{A \Omega^2}{A^2 + 2 \Omega^2}
\right)\tau_{\rm p} \right\} ~.
\end{equation}
In Eq.~(\ref{H41}) we bound the absolute value of the expression in
the curly  brackets by using $|\sin b\tau_{\rm p}/b| \le \tau_{\rm p}$ and
$|\sin b\tau_{\rm p} |,
|\cos b\tau_{\rm p} | \le 1$. With Eq.~(\ref{H42}) this leads to an upper bound
of the form
\[
{\rm e}^{- \gamma t}]{\rm e}^{-N_{\rm e}}~.
\eta'~{\rm const} ~[(\alpha + \beta \tau_{\rm p})
{\rm e}^{- \gamma \tau_{\rm p}}]~{\rm
e}^{-N_{\rm e}}~. \]
The maximum of the expression in the square brackets is taken for 
$\tau_{\rm p}=\tau_{\rm m}=\frac{1}{\gamma}-
\frac{\alpha}{\beta}$ if $\tau_{\rm m} 
\ge 0$, otherwise for $0$. One can check graphically that $\tau_{\rm m} 
< 0$ for $ \eta' \le 1/3$
and hence the maximum is taken at $\tau_{\rm p} = 0$. An upper bound,
valid for
all $A$ and $\Omega$, is then found graphically,
with $N_{\rm e} = N_{\rm d}/\eta'$ and 
for $\eta' \le 1/3$, as 
\begin{equation}\label{H42a}
|c_2 {\rm e}^{p_2 \tau_{\rm p}} + c_3 {\rm e}^{p_3 \tau_{\rm p}}| \le 
0.81 \eta' ~{\rm e}^{-N_{\rm d}/\eta'}~. 
\end{equation}
Together with the bound for $c_1 {\rm e}^{p_1 \tau_{\rm p}}$ in
Eq.~(\ref{32e}) this gives, for $\eta' \le 1/3$, 
\begin{equation}\label{H42b}
P^{\rm d}_0 (\tau_{\rm p} ; |1 \rangle) \le \left\{ \frac{1}{1 - 0.45
  \eta'}~+ 0.81 \eta' {\rm e}^{-\left( \frac{1}{\eta'}~- 1 \right) N_{\rm d}}
\right\} {\rm e}^{-N_{\rm d}}~.
\end{equation}
The expression in the curly brackets can be evaluated graphically, and
this leads, for $N_{\rm d} \ge 1$ and $\eta' \le 1/3$, to
\begin{equation}\label{H43}
P^{\rm d}_0 (\tau_{\rm p} ; |1 \rangle) \le \left( 1
+~\frac{2}{3}\eta' \right) {\rm e}^{-
  N_{\rm d} (\tau_{\rm p}; |1 \rangle)}~.
\end{equation}

In a similar way one can also obtain a lower bound for $P^{\rm d}_0$, but it
is quicker to use Jensen's inequality \cite{Breiman} which gives
\[
\sum^\infty_{N=0} (1-\eta')^N p_N \ge (1-\eta')^{\bar{N}}~.
\]
With $\bar{N} \equiv N_{\rm e}$ Eq.~(\ref{26}) then yields 
\[
(1 - \eta')^{N_{\rm e}} = {\rm e}^{\mathop{\rm ln}(1-\eta')N_{\rm e}}
\le P^{\rm d}_0~.
\]
Since $|{\rm ln} (1-\eta')| \le \eta' +~\frac{1}{2}~\eta'^2$ one
finally obtains, with $N_{\rm d} \equiv N_{\rm d} (\tau_{\rm p}; |1
\rangle)$ and $P^{\rm d}_0
\equiv P^{\rm d}_0 (\tau_{\rm p}; |1 \rangle)$,
\begin{equation}\label{H44}
{\rm e}^{-(1 + \eta'/2)N_{\rm d}} \le
(1 - \eta')^{N_{\rm d}/\eta'} \le P^{\rm d}_0 \le (1
+~\frac{2}{3}~\eta') {\rm e}^{-N_{\rm d}}
\end{equation}
where the l.h.s. always holds and the r.h.s. for $N_{\rm d} \ge 1$ and $0
\le \eta' \le 1/3$.

Eq.~(\ref{H44}) not only shows that $P^0_{\rm d} \approx 
{\rm e}^{-N_{\rm d}}$ 
is indeed
an excellent approximation for small detector efficiencies, as already
suggested in Section 3, but  also gives strict error bounds.

\section{\bf Conclusions}
A short laser pulse driving the 1--3 transition brings about a
reduction of a state of the 1--2 system in two ways as follows. If one
or more fluorescence photons are emitted then the system ends up in
$|1 \rangle$, after a transient time for the decay of level 3. If no photons
are emitted then the reduced (conditional) time development pushes the
state towards $|2 \rangle$. Its tiny $|1\rangle$ component can be
estimated from above by
\[
|\langle 1|\psi\rangle|^2 {\rm e}^{-N_{\rm e}}~.
\]
Thus the reduction of $|\psi \rangle$ to $|2\rangle$ is excellent if
the parameters of the pulse are such that $N_{\rm e} \ge 5$, i.e.
if the average number of photons emitted by an atom initially in
the ground state is 5 or more.
If one measures with a detector which does not detect all photons 
$(\eta' < 1$, e.g. solid angle less than $4 \pi)$, then one has a problem
for single systems. 

Detecting no fluorescence photons during a pulse may be due either to
no emission at all or to missing the emitted photons. As a consequence
a system with initial state $|\psi \rangle$ for which no fluorescence
photons are detected during a pulse is in a mixture of $|1 \rangle$ and
$|2 \rangle$, namely
\begin{equation}
|\langle 2|\psi \rangle|^2~|2\rangle \langle 2|+
|\langle 1|\psi \rangle|^2P_0^{\rm d}(\tau_{\rm p};|1\rangle)~|1\rangle
\langle 1|~. \end{equation}
$P^{\rm d}_0 (\tau_{\rm p} ; |1\rangle)$ is the probability of {\em
detecting} no fluorescence photons for a pulse, with the system
initially in the ground state. We have expressed $P^{\rm d}_0$ in terms
of the easily measurable quantity $N_{\rm d} (\tau_{\rm p} ; |1
\rangle)$, the average number of detected photons per system, with
initial state $|1 \rangle$. $N_{\rm d}$ is simply determined from the
fluorescence rate of an ensemble of systems initially in the ground
state. For small efficiency $\eta'$ we have shown that
\[
P^{\rm d}_0 = {\rm e}^{-N_{\rm d}}
\]
and, more precisely, for $\eta' \le 1/3$ and $N_{\rm d} \ge 1$, 
\[
{\rm e}^{-(1+\eta'/2)N_{\rm d}} \le P^{\rm d}_0 \le (1 
+~\frac{2}{3}~\eta'){\rm e}^{-N_{\rm d}}~.
\]
Thus for increasing $N_{\rm d}$ the ``non-reduced part'' becomes very 
small, e.g. for $N_{\rm d} = 3$ it is already less than 5 \%.

Depending on the desired quality of the reduction one can adjust the
parameters of the laser pulse in a simple way to give the required
value of $N_{\rm d}$.

If one considers not a single system but an ensemble of atoms the
detector plays no role. After the pulse the
state is reduced to a mixture of $|1 \rangle$ and $|2\rangle$ regardless
of whether all fluorescence photons are actually observed or not. This
can also be seen by the use of Bloch equations \cite{Block,Fre}.

If the 1-2 transition
is driven by an additional interaction during the short laser pulse then
complications arise and modifications to the projection-postulate
result as obtained above have to be incorporated 
\cite{BeHe,BeHeSo}.\\[0.5cm]

{\bf Acknowledgments}. We would like to thank D.G. Sondermann for
stimulating discussions.

\newpage

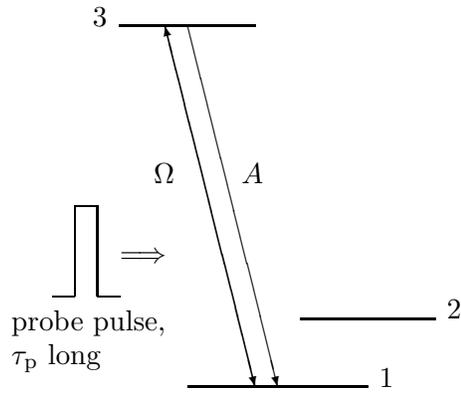
\begin{figure}
\unitlength 0.6cm
\begin{picture}(18,11)
\thicklines
\put(6,9.5) {\line(1,0){3}}
\put(10,3) {\line(1,0){3}}
\put(7.5,1.5) {\line(1,0){4}}
\thinlines
\put(4.5,3.5) {\line(1,0){0.5}}
\put(5.5,3.5) {\line(1,0){0.5}}
\put(5,5.5) {\line(1,0){0.5}}
\put(5,3.5) {\line(0,1){2}}
\put(5.5,3.5) {\line(0,1){2}}
\put(7,9.5) {\vector(1,-4){2}}
\put(7.5,9.5) {\vector(1,-4){2}}
\put(9,1.5) {\vector(-1,4){2}}
\put (5.4,9.5){3}
\put (6,4.25){$\Longrightarrow $}
\put (6.75,6){$\Omega$}
\put (8.7,6){$A$}
\put (3.6,2.75){probe pulse,}
\put (3.6,2){$\tau_{\rm p}$ long}
\put (13.25,3){2}
\put (11.75,1.5){1}
\end{picture}

{\caption{ V system with level 2 and auxiliary level 3
with Einstein coefficient $A$. $\Omega$ is the Rabi
frequencies of the probe pulse of length $\tau_{\rm p}$.}}
\end{figure}

\begin{figure}
\setlength{\unitlength}{0.240900pt}
\ifx\plotpoint\undefined\newsavebox{\plotpoint}\fi
\sbox{\plotpoint}{\rule[-0.200pt]{0.400pt}{0.400pt}}%
\begin{picture}(1200,900)(0,0)
\font\gnuplot=cmr10 at 10pt
\gnuplot
\sbox{\plotpoint}{\rule[-0.200pt]{0.400pt}{0.400pt}}%
\put(981.0,113.0){\rule[-0.200pt]{0.400pt}{184.048pt}}
\put(176.0,363.0){\rule[-0.200pt]{4.818pt}{0.400pt}}
\put(154,363){\makebox(0,0)[r]{$\eta'=1$}}
\put(1116.0,363.0){\rule[-0.200pt]{4.818pt}{0.400pt}}
\put(176.0,475.0){\rule[-0.200pt]{4.818pt}{0.400pt}}
\put(154,475){\makebox(0,0)[r]{$\eta'=0$}}
\put(1116.0,475.0){\rule[-0.200pt]{4.818pt}{0.400pt}}
\put(612.0,113.0){\rule[-0.200pt]{0.400pt}{4.818pt}}
\put(612,68){\makebox(0,0){$p_{\rm infl}$}}
\put(612.0,857.0){\rule[-0.200pt]{0.400pt}{4.818pt}}
\put(981.0,113.0){\rule[-0.200pt]{0.400pt}{4.818pt}}
\put(981,68){\makebox(0,0){0}}
\put(981.0,857.0){\rule[-0.200pt]{0.400pt}{4.818pt}}
\put(176.0,113.0){\rule[-0.200pt]{231.264pt}{0.400pt}}
\put(1136.0,113.0){\rule[-0.200pt]{0.400pt}{184.048pt}}
\put(176.0,877.0){\rule[-0.200pt]{231.264pt}{0.400pt}}
\put(656,23){\makebox(0,0){$p$}}
\put(176.0,113.0){\rule[-0.200pt]{0.400pt}{184.048pt}}
\multiput(763.59,113.00)(0.477,0.710){7}{\rule{0.115pt}{0.660pt}}
\multiput(762.17,113.00)(5.000,5.630){2}{\rule{0.400pt}{0.330pt}}
\multiput(768.59,120.00)(0.489,0.902){15}{\rule{0.118pt}{0.811pt}}
\multiput(767.17,120.00)(9.000,14.316){2}{\rule{0.400pt}{0.406pt}}
\multiput(777.58,136.00)(0.491,0.808){17}{\rule{0.118pt}{0.740pt}}
\multiput(776.17,136.00)(10.000,14.464){2}{\rule{0.400pt}{0.370pt}}
\multiput(787.58,152.00)(0.491,0.808){17}{\rule{0.118pt}{0.740pt}}
\multiput(786.17,152.00)(10.000,14.464){2}{\rule{0.400pt}{0.370pt}}
\multiput(797.59,168.00)(0.489,0.902){15}{\rule{0.118pt}{0.811pt}}
\multiput(796.17,168.00)(9.000,14.316){2}{\rule{0.400pt}{0.406pt}}
\multiput(806.58,184.00)(0.491,0.808){17}{\rule{0.118pt}{0.740pt}}
\multiput(805.17,184.00)(10.000,14.464){2}{\rule{0.400pt}{0.370pt}}
\multiput(816.58,200.00)(0.491,0.860){17}{\rule{0.118pt}{0.780pt}}
\multiput(815.17,200.00)(10.000,15.381){2}{\rule{0.400pt}{0.390pt}}
\multiput(826.59,217.00)(0.489,0.902){15}{\rule{0.118pt}{0.811pt}}
\multiput(825.17,217.00)(9.000,14.316){2}{\rule{0.400pt}{0.406pt}}
\multiput(835.58,233.00)(0.491,0.808){17}{\rule{0.118pt}{0.740pt}}
\multiput(834.17,233.00)(10.000,14.464){2}{\rule{0.400pt}{0.370pt}}
\multiput(845.58,249.00)(0.491,0.808){17}{\rule{0.118pt}{0.740pt}}
\multiput(844.17,249.00)(10.000,14.464){2}{\rule{0.400pt}{0.370pt}}
\multiput(855.59,265.00)(0.489,0.902){15}{\rule{0.118pt}{0.811pt}}
\multiput(854.17,265.00)(9.000,14.316){2}{\rule{0.400pt}{0.406pt}}
\multiput(864.58,281.00)(0.491,0.808){17}{\rule{0.118pt}{0.740pt}}
\multiput(863.17,281.00)(10.000,14.464){2}{\rule{0.400pt}{0.370pt}}
\multiput(874.58,297.00)(0.491,0.808){17}{\rule{0.118pt}{0.740pt}}
\multiput(873.17,297.00)(10.000,14.464){2}{\rule{0.400pt}{0.370pt}}
\multiput(884.58,313.00)(0.491,0.860){17}{\rule{0.118pt}{0.780pt}}
\multiput(883.17,313.00)(10.000,15.381){2}{\rule{0.400pt}{0.390pt}}
\multiput(894.59,330.00)(0.489,0.902){15}{\rule{0.118pt}{0.811pt}}
\multiput(893.17,330.00)(9.000,14.316){2}{\rule{0.400pt}{0.406pt}}
\multiput(903.58,346.00)(0.491,0.808){17}{\rule{0.118pt}{0.740pt}}
\multiput(902.17,346.00)(10.000,14.464){2}{\rule{0.400pt}{0.370pt}}
\multiput(913.58,362.00)(0.491,0.808){17}{\rule{0.118pt}{0.740pt}}
\multiput(912.17,362.00)(10.000,14.464){2}{\rule{0.400pt}{0.370pt}}
\multiput(923.59,378.00)(0.489,0.902){15}{\rule{0.118pt}{0.811pt}}
\multiput(922.17,378.00)(9.000,14.316){2}{\rule{0.400pt}{0.406pt}}
\multiput(932.58,394.00)(0.491,0.808){17}{\rule{0.118pt}{0.740pt}}
\multiput(931.17,394.00)(10.000,14.464){2}{\rule{0.400pt}{0.370pt}}
\multiput(942.58,410.00)(0.491,0.808){17}{\rule{0.118pt}{0.740pt}}
\multiput(941.17,410.00)(10.000,14.464){2}{\rule{0.400pt}{0.370pt}}
\multiput(952.59,426.00)(0.489,0.961){15}{\rule{0.118pt}{0.856pt}}
\multiput(951.17,426.00)(9.000,15.224){2}{\rule{0.400pt}{0.428pt}}
\multiput(961.58,443.00)(0.491,0.808){17}{\rule{0.118pt}{0.740pt}}
\multiput(960.17,443.00)(10.000,14.464){2}{\rule{0.400pt}{0.370pt}}
\multiput(971.58,459.00)(0.491,0.808){17}{\rule{0.118pt}{0.740pt}}
\multiput(970.17,459.00)(10.000,14.464){2}{\rule{0.400pt}{0.370pt}}
\multiput(981.58,475.00)(0.491,0.808){17}{\rule{0.118pt}{0.740pt}}
\multiput(980.17,475.00)(10.000,14.464){2}{\rule{0.400pt}{0.370pt}}
\multiput(991.59,491.00)(0.489,0.902){15}{\rule{0.118pt}{0.811pt}}
\multiput(990.17,491.00)(9.000,14.316){2}{\rule{0.400pt}{0.406pt}}
\multiput(1000.58,507.00)(0.491,0.808){17}{\rule{0.118pt}{0.740pt}}
\multiput(999.17,507.00)(10.000,14.464){2}{\rule{0.400pt}{0.370pt}}
\multiput(1010.58,523.00)(0.491,0.808){17}{\rule{0.118pt}{0.740pt}}
\multiput(1009.17,523.00)(10.000,14.464){2}{\rule{0.400pt}{0.370pt}}
\multiput(1020.59,539.00)(0.489,0.902){15}{\rule{0.118pt}{0.811pt}}
\multiput(1019.17,539.00)(9.000,14.316){2}{\rule{0.400pt}{0.406pt}}
\multiput(1029.58,555.00)(0.491,0.860){17}{\rule{0.118pt}{0.780pt}}
\multiput(1028.17,555.00)(10.000,15.381){2}{\rule{0.400pt}{0.390pt}}
\multiput(1039.58,572.00)(0.491,0.808){17}{\rule{0.118pt}{0.740pt}}
\multiput(1038.17,572.00)(10.000,14.464){2}{\rule{0.400pt}{0.370pt}}
\multiput(1049.59,588.00)(0.489,0.902){15}{\rule{0.118pt}{0.811pt}}
\multiput(1048.17,588.00)(9.000,14.316){2}{\rule{0.400pt}{0.406pt}}
\multiput(1058.58,604.00)(0.491,0.808){17}{\rule{0.118pt}{0.740pt}}
\multiput(1057.17,604.00)(10.000,14.464){2}{\rule{0.400pt}{0.370pt}}
\multiput(1068.58,620.00)(0.491,0.808){17}{\rule{0.118pt}{0.740pt}}
\multiput(1067.17,620.00)(10.000,14.464){2}{\rule{0.400pt}{0.370pt}}
\multiput(1078.58,636.00)(0.491,0.808){17}{\rule{0.118pt}{0.740pt}}
\multiput(1077.17,636.00)(10.000,14.464){2}{\rule{0.400pt}{0.370pt}}
\multiput(1088.59,652.00)(0.489,0.902){15}{\rule{0.118pt}{0.811pt}}
\multiput(1087.17,652.00)(9.000,14.316){2}{\rule{0.400pt}{0.406pt}}
\multiput(1097.58,668.00)(0.491,0.860){17}{\rule{0.118pt}{0.780pt}}
\multiput(1096.17,668.00)(10.000,15.381){2}{\rule{0.400pt}{0.390pt}}
\multiput(1107.58,685.00)(0.491,0.808){17}{\rule{0.118pt}{0.740pt}}
\multiput(1106.17,685.00)(10.000,14.464){2}{\rule{0.400pt}{0.370pt}}
\multiput(1117.59,701.00)(0.489,0.902){15}{\rule{0.118pt}{0.811pt}}
\multiput(1116.17,701.00)(9.000,14.316){2}{\rule{0.400pt}{0.406pt}}
\multiput(1126.58,717.00)(0.491,0.808){17}{\rule{0.118pt}{0.740pt}}
\multiput(1125.17,717.00)(10.000,14.464){2}{\rule{0.400pt}{0.370pt}}
\sbox{\plotpoint}{\rule[-0.400pt]{0.800pt}{0.800pt}}%
\put(176,115){\usebox{\plotpoint}}
\multiput(177.40,115.00)(0.514,1.211){13}{\rule{0.124pt}{2.040pt}}
\multiput(174.34,115.00)(10.000,18.766){2}{\rule{0.800pt}{1.020pt}}
\multiput(187.40,138.00)(0.516,1.305){11}{\rule{0.124pt}{2.156pt}}
\multiput(184.34,138.00)(9.000,17.526){2}{\rule{0.800pt}{1.078pt}}
\multiput(196.40,160.00)(0.514,1.100){13}{\rule{0.124pt}{1.880pt}}
\multiput(193.34,160.00)(10.000,17.098){2}{\rule{0.800pt}{0.940pt}}
\multiput(206.40,181.00)(0.514,1.044){13}{\rule{0.124pt}{1.800pt}}
\multiput(203.34,181.00)(10.000,16.264){2}{\rule{0.800pt}{0.900pt}}
\multiput(216.40,201.00)(0.516,1.116){11}{\rule{0.124pt}{1.889pt}}
\multiput(213.34,201.00)(9.000,15.080){2}{\rule{0.800pt}{0.944pt}}
\multiput(225.40,220.00)(0.514,0.877){13}{\rule{0.124pt}{1.560pt}}
\multiput(222.34,220.00)(10.000,13.762){2}{\rule{0.800pt}{0.780pt}}
\multiput(235.40,237.00)(0.514,0.877){13}{\rule{0.124pt}{1.560pt}}
\multiput(232.34,237.00)(10.000,13.762){2}{\rule{0.800pt}{0.780pt}}
\multiput(245.40,254.00)(0.514,0.766){13}{\rule{0.124pt}{1.400pt}}
\multiput(242.34,254.00)(10.000,12.094){2}{\rule{0.800pt}{0.700pt}}
\multiput(255.40,269.00)(0.516,0.863){11}{\rule{0.124pt}{1.533pt}}
\multiput(252.34,269.00)(9.000,11.817){2}{\rule{0.800pt}{0.767pt}}
\multiput(264.40,284.00)(0.514,0.654){13}{\rule{0.124pt}{1.240pt}}
\multiput(261.34,284.00)(10.000,10.426){2}{\rule{0.800pt}{0.620pt}}
\multiput(274.40,297.00)(0.514,0.654){13}{\rule{0.124pt}{1.240pt}}
\multiput(271.34,297.00)(10.000,10.426){2}{\rule{0.800pt}{0.620pt}}
\multiput(284.40,310.00)(0.516,0.611){11}{\rule{0.124pt}{1.178pt}}
\multiput(281.34,310.00)(9.000,8.555){2}{\rule{0.800pt}{0.589pt}}
\multiput(293.40,321.00)(0.514,0.543){13}{\rule{0.124pt}{1.080pt}}
\multiput(290.34,321.00)(10.000,8.758){2}{\rule{0.800pt}{0.540pt}}
\multiput(302.00,333.40)(0.487,0.514){13}{\rule{1.000pt}{0.124pt}}
\multiput(302.00,330.34)(7.924,10.000){2}{\rule{0.500pt}{0.800pt}}
\multiput(312.00,343.40)(0.485,0.516){11}{\rule{1.000pt}{0.124pt}}
\multiput(312.00,340.34)(6.924,9.000){2}{\rule{0.500pt}{0.800pt}}
\multiput(321.00,352.40)(0.627,0.520){9}{\rule{1.200pt}{0.125pt}}
\multiput(321.00,349.34)(7.509,8.000){2}{\rule{0.600pt}{0.800pt}}
\multiput(331.00,360.40)(0.738,0.526){7}{\rule{1.343pt}{0.127pt}}
\multiput(331.00,357.34)(7.213,7.000){2}{\rule{0.671pt}{0.800pt}}
\multiput(341.00,367.40)(0.738,0.526){7}{\rule{1.343pt}{0.127pt}}
\multiput(341.00,364.34)(7.213,7.000){2}{\rule{0.671pt}{0.800pt}}
\multiput(351.00,374.39)(0.797,0.536){5}{\rule{1.400pt}{0.129pt}}
\multiput(351.00,371.34)(6.094,6.000){2}{\rule{0.700pt}{0.800pt}}
\multiput(360.00,380.38)(1.264,0.560){3}{\rule{1.800pt}{0.135pt}}
\multiput(360.00,377.34)(6.264,5.000){2}{\rule{0.900pt}{0.800pt}}
\multiput(370.00,385.38)(1.264,0.560){3}{\rule{1.800pt}{0.135pt}}
\multiput(370.00,382.34)(6.264,5.000){2}{\rule{0.900pt}{0.800pt}}
\put(380,388.84){\rule{2.168pt}{0.800pt}}
\multiput(380.00,387.34)(4.500,3.000){2}{\rule{1.084pt}{0.800pt}}
\put(389,392.34){\rule{2.200pt}{0.800pt}}
\multiput(389.00,390.34)(5.434,4.000){2}{\rule{1.100pt}{0.800pt}}
\put(399,395.34){\rule{2.409pt}{0.800pt}}
\multiput(399.00,394.34)(5.000,2.000){2}{\rule{1.204pt}{0.800pt}}
\put(409,397.34){\rule{2.168pt}{0.800pt}}
\multiput(409.00,396.34)(4.500,2.000){2}{\rule{1.084pt}{0.800pt}}
\put(418,399.34){\rule{2.409pt}{0.800pt}}
\multiput(418.00,398.34)(5.000,2.000){2}{\rule{1.204pt}{0.800pt}}
\put(428,400.84){\rule{2.409pt}{0.800pt}}
\multiput(428.00,400.34)(5.000,1.000){2}{\rule{1.204pt}{0.800pt}}
\put(467,400.84){\rule{2.409pt}{0.800pt}}
\multiput(467.00,401.34)(5.000,-1.000){2}{\rule{1.204pt}{0.800pt}}
\put(477,399.84){\rule{2.168pt}{0.800pt}}
\multiput(477.00,400.34)(4.500,-1.000){2}{\rule{1.084pt}{0.800pt}}
\put(486,398.34){\rule{2.409pt}{0.800pt}}
\multiput(486.00,399.34)(5.000,-2.000){2}{\rule{1.204pt}{0.800pt}}
\put(496,396.34){\rule{2.409pt}{0.800pt}}
\multiput(496.00,397.34)(5.000,-2.000){2}{\rule{1.204pt}{0.800pt}}
\put(506,394.34){\rule{2.168pt}{0.800pt}}
\multiput(506.00,395.34)(4.500,-2.000){2}{\rule{1.084pt}{0.800pt}}
\put(515,392.34){\rule{2.409pt}{0.800pt}}
\multiput(515.00,393.34)(5.000,-2.000){2}{\rule{1.204pt}{0.800pt}}
\put(525,389.84){\rule{2.409pt}{0.800pt}}
\multiput(525.00,391.34)(5.000,-3.000){2}{\rule{1.204pt}{0.800pt}}
\put(535,386.84){\rule{2.168pt}{0.800pt}}
\multiput(535.00,388.34)(4.500,-3.000){2}{\rule{1.084pt}{0.800pt}}
\put(544,383.84){\rule{2.409pt}{0.800pt}}
\multiput(544.00,385.34)(5.000,-3.000){2}{\rule{1.204pt}{0.800pt}}
\put(554,380.84){\rule{2.409pt}{0.800pt}}
\multiput(554.00,382.34)(5.000,-3.000){2}{\rule{1.204pt}{0.800pt}}
\put(564,377.34){\rule{2.200pt}{0.800pt}}
\multiput(564.00,379.34)(5.434,-4.000){2}{\rule{1.100pt}{0.800pt}}
\put(574,373.84){\rule{2.168pt}{0.800pt}}
\multiput(574.00,375.34)(4.500,-3.000){2}{\rule{1.084pt}{0.800pt}}
\put(583,370.34){\rule{2.200pt}{0.800pt}}
\multiput(583.00,372.34)(5.434,-4.000){2}{\rule{1.100pt}{0.800pt}}
\put(593,366.84){\rule{2.409pt}{0.800pt}}
\multiput(593.00,368.34)(5.000,-3.000){2}{\rule{1.204pt}{0.800pt}}
\put(603,363.34){\rule{2.000pt}{0.800pt}}
\multiput(603.00,365.34)(4.849,-4.000){2}{\rule{1.000pt}{0.800pt}}
\put(612,359.34){\rule{2.200pt}{0.800pt}}
\multiput(612.00,361.34)(5.434,-4.000){2}{\rule{1.100pt}{0.800pt}}
\put(622,355.84){\rule{2.409pt}{0.800pt}}
\multiput(622.00,357.34)(5.000,-3.000){2}{\rule{1.204pt}{0.800pt}}
\put(632,352.34){\rule{2.000pt}{0.800pt}}
\multiput(632.00,354.34)(4.849,-4.000){2}{\rule{1.000pt}{0.800pt}}
\put(641,348.84){\rule{2.409pt}{0.800pt}}
\multiput(641.00,350.34)(5.000,-3.000){2}{\rule{1.204pt}{0.800pt}}
\put(651,345.34){\rule{2.200pt}{0.800pt}}
\multiput(651.00,347.34)(5.434,-4.000){2}{\rule{1.100pt}{0.800pt}}
\put(661,341.84){\rule{2.409pt}{0.800pt}}
\multiput(661.00,343.34)(5.000,-3.000){2}{\rule{1.204pt}{0.800pt}}
\put(671,338.84){\rule{2.168pt}{0.800pt}}
\multiput(671.00,340.34)(4.500,-3.000){2}{\rule{1.084pt}{0.800pt}}
\put(680,335.84){\rule{2.409pt}{0.800pt}}
\multiput(680.00,337.34)(5.000,-3.000){2}{\rule{1.204pt}{0.800pt}}
\put(690,332.84){\rule{2.409pt}{0.800pt}}
\multiput(690.00,334.34)(5.000,-3.000){2}{\rule{1.204pt}{0.800pt}}
\put(700,330.34){\rule{2.168pt}{0.800pt}}
\multiput(700.00,331.34)(4.500,-2.000){2}{\rule{1.084pt}{0.800pt}}
\put(709,328.34){\rule{2.409pt}{0.800pt}}
\multiput(709.00,329.34)(5.000,-2.000){2}{\rule{1.204pt}{0.800pt}}
\put(719,326.34){\rule{2.409pt}{0.800pt}}
\multiput(719.00,327.34)(5.000,-2.000){2}{\rule{1.204pt}{0.800pt}}
\put(729,324.34){\rule{2.168pt}{0.800pt}}
\multiput(729.00,325.34)(4.500,-2.000){2}{\rule{1.084pt}{0.800pt}}
\put(738,322.84){\rule{2.409pt}{0.800pt}}
\multiput(738.00,323.34)(5.000,-1.000){2}{\rule{1.204pt}{0.800pt}}
\put(438.0,403.0){\rule[-0.400pt]{6.986pt}{0.800pt}}
\put(758,321.84){\rule{2.409pt}{0.800pt}}
\multiput(758.00,322.34)(5.000,-1.000){2}{\rule{1.204pt}{0.800pt}}
\put(748.0,324.0){\rule[-0.400pt]{2.409pt}{0.800pt}}
\put(777,321.84){\rule{2.409pt}{0.800pt}}
\multiput(777.00,321.34)(5.000,1.000){2}{\rule{1.204pt}{0.800pt}}
\put(787,322.84){\rule{2.409pt}{0.800pt}}
\multiput(787.00,322.34)(5.000,1.000){2}{\rule{1.204pt}{0.800pt}}
\put(797,323.84){\rule{2.168pt}{0.800pt}}
\multiput(797.00,323.34)(4.500,1.000){2}{\rule{1.084pt}{0.800pt}}
\put(806,325.84){\rule{2.409pt}{0.800pt}}
\multiput(806.00,324.34)(5.000,3.000){2}{\rule{1.204pt}{0.800pt}}
\put(816,328.34){\rule{2.409pt}{0.800pt}}
\multiput(816.00,327.34)(5.000,2.000){2}{\rule{1.204pt}{0.800pt}}
\put(826,330.84){\rule{2.168pt}{0.800pt}}
\multiput(826.00,329.34)(4.500,3.000){2}{\rule{1.084pt}{0.800pt}}
\put(835,334.34){\rule{2.200pt}{0.800pt}}
\multiput(835.00,332.34)(5.434,4.000){2}{\rule{1.100pt}{0.800pt}}
\multiput(845.00,339.38)(1.264,0.560){3}{\rule{1.800pt}{0.135pt}}
\multiput(845.00,336.34)(6.264,5.000){2}{\rule{0.900pt}{0.800pt}}
\multiput(855.00,344.38)(1.096,0.560){3}{\rule{1.640pt}{0.135pt}}
\multiput(855.00,341.34)(5.596,5.000){2}{\rule{0.820pt}{0.800pt}}
\multiput(864.00,349.39)(0.909,0.536){5}{\rule{1.533pt}{0.129pt}}
\multiput(864.00,346.34)(6.817,6.000){2}{\rule{0.767pt}{0.800pt}}
\multiput(874.00,355.40)(0.738,0.526){7}{\rule{1.343pt}{0.127pt}}
\multiput(874.00,352.34)(7.213,7.000){2}{\rule{0.671pt}{0.800pt}}
\multiput(884.00,362.40)(0.738,0.526){7}{\rule{1.343pt}{0.127pt}}
\multiput(884.00,359.34)(7.213,7.000){2}{\rule{0.671pt}{0.800pt}}
\multiput(894.00,369.40)(0.485,0.516){11}{\rule{1.000pt}{0.124pt}}
\multiput(894.00,366.34)(6.924,9.000){2}{\rule{0.500pt}{0.800pt}}
\multiput(903.00,378.40)(0.548,0.516){11}{\rule{1.089pt}{0.124pt}}
\multiput(903.00,375.34)(7.740,9.000){2}{\rule{0.544pt}{0.800pt}}
\multiput(913.00,387.40)(0.487,0.514){13}{\rule{1.000pt}{0.124pt}}
\multiput(913.00,384.34)(7.924,10.000){2}{\rule{0.500pt}{0.800pt}}
\multiput(924.40,396.00)(0.516,0.611){11}{\rule{0.124pt}{1.178pt}}
\multiput(921.34,396.00)(9.000,8.555){2}{\rule{0.800pt}{0.589pt}}
\multiput(933.40,407.00)(0.514,0.543){13}{\rule{0.124pt}{1.080pt}}
\multiput(930.34,407.00)(10.000,8.758){2}{\rule{0.800pt}{0.540pt}}
\multiput(943.40,418.00)(0.514,0.654){13}{\rule{0.124pt}{1.240pt}}
\multiput(940.34,418.00)(10.000,10.426){2}{\rule{0.800pt}{0.620pt}}
\multiput(953.40,431.00)(0.516,0.800){11}{\rule{0.124pt}{1.444pt}}
\multiput(950.34,431.00)(9.000,11.002){2}{\rule{0.800pt}{0.722pt}}
\multiput(962.40,445.00)(0.514,0.710){13}{\rule{0.124pt}{1.320pt}}
\multiput(959.34,445.00)(10.000,11.260){2}{\rule{0.800pt}{0.660pt}}
\multiput(972.40,459.00)(0.514,0.821){13}{\rule{0.124pt}{1.480pt}}
\multiput(969.34,459.00)(10.000,12.928){2}{\rule{0.800pt}{0.740pt}}
\multiput(982.40,475.00)(0.514,0.821){13}{\rule{0.124pt}{1.480pt}}
\multiput(979.34,475.00)(10.000,12.928){2}{\rule{0.800pt}{0.740pt}}
\multiput(992.40,491.00)(0.516,1.053){11}{\rule{0.124pt}{1.800pt}}
\multiput(989.34,491.00)(9.000,14.264){2}{\rule{0.800pt}{0.900pt}}
\multiput(1001.40,509.00)(0.514,0.988){13}{\rule{0.124pt}{1.720pt}}
\multiput(998.34,509.00)(10.000,15.430){2}{\rule{0.800pt}{0.860pt}}
\multiput(1011.40,528.00)(0.514,1.044){13}{\rule{0.124pt}{1.800pt}}
\multiput(1008.34,528.00)(10.000,16.264){2}{\rule{0.800pt}{0.900pt}}
\multiput(1021.40,548.00)(0.516,1.242){11}{\rule{0.124pt}{2.067pt}}
\multiput(1018.34,548.00)(9.000,16.711){2}{\rule{0.800pt}{1.033pt}}
\multiput(1030.40,569.00)(0.514,1.155){13}{\rule{0.124pt}{1.960pt}}
\multiput(1027.34,569.00)(10.000,17.932){2}{\rule{0.800pt}{0.980pt}}
\multiput(1040.40,591.00)(0.514,1.267){13}{\rule{0.124pt}{2.120pt}}
\multiput(1037.34,591.00)(10.000,19.600){2}{\rule{0.800pt}{1.060pt}}
\multiput(1050.40,615.00)(0.516,1.431){11}{\rule{0.124pt}{2.333pt}}
\multiput(1047.34,615.00)(9.000,19.157){2}{\rule{0.800pt}{1.167pt}}
\multiput(1059.40,639.00)(0.514,1.378){13}{\rule{0.124pt}{2.280pt}}
\multiput(1056.34,639.00)(10.000,21.268){2}{\rule{0.800pt}{1.140pt}}
\multiput(1069.40,665.00)(0.514,1.489){13}{\rule{0.124pt}{2.440pt}}
\multiput(1066.34,665.00)(10.000,22.936){2}{\rule{0.800pt}{1.220pt}}
\multiput(1079.40,693.00)(0.514,1.489){13}{\rule{0.124pt}{2.440pt}}
\multiput(1076.34,693.00)(10.000,22.936){2}{\rule{0.800pt}{1.220pt}}
\multiput(1089.40,721.00)(0.516,1.810){11}{\rule{0.124pt}{2.867pt}}
\multiput(1086.34,721.00)(9.000,24.050){2}{\rule{0.800pt}{1.433pt}}
\multiput(1098.40,751.00)(0.514,1.656){13}{\rule{0.124pt}{2.680pt}}
\multiput(1095.34,751.00)(10.000,25.438){2}{\rule{0.800pt}{1.340pt}}
\multiput(1108.40,782.00)(0.514,1.768){13}{\rule{0.124pt}{2.840pt}}
\multiput(1105.34,782.00)(10.000,27.105){2}{\rule{0.800pt}{1.420pt}}
\multiput(1118.40,815.00)(0.516,2.062){11}{\rule{0.124pt}{3.222pt}}
\multiput(1115.34,815.00)(9.000,27.312){2}{\rule{0.800pt}{1.611pt}}
\multiput(1127.40,849.00)(0.520,1.943){9}{\rule{0.125pt}{3.000pt}}
\multiput(1124.34,849.00)(8.000,21.773){2}{\rule{0.800pt}{1.500pt}}
\put(768.0,323.0){\rule[-0.400pt]{2.168pt}{0.800pt}}
\sbox{\plotpoint}{\rule[-0.500pt]{1.000pt}{1.000pt}}%
\put(176,363){\usebox{\plotpoint}}
\put(176.00,363.00){\usebox{\plotpoint}}
\multiput(186,363)(20.756,0.000){0}{\usebox{\plotpoint}}
\put(196.76,363.00){\usebox{\plotpoint}}
\multiput(205,363)(20.756,0.000){0}{\usebox{\plotpoint}}
\put(217.51,363.00){\usebox{\plotpoint}}
\multiput(224,363)(20.756,0.000){0}{\usebox{\plotpoint}}
\put(238.27,363.00){\usebox{\plotpoint}}
\multiput(244,363)(20.756,0.000){0}{\usebox{\plotpoint}}
\put(259.02,363.00){\usebox{\plotpoint}}
\multiput(263,363)(20.756,0.000){0}{\usebox{\plotpoint}}
\put(279.78,363.00){\usebox{\plotpoint}}
\multiput(283,363)(20.756,0.000){0}{\usebox{\plotpoint}}
\put(300.53,363.00){\usebox{\plotpoint}}
\multiput(302,363)(20.756,0.000){0}{\usebox{\plotpoint}}
\multiput(312,363)(20.756,0.000){0}{\usebox{\plotpoint}}
\put(321.29,363.00){\usebox{\plotpoint}}
\multiput(331,363)(20.756,0.000){0}{\usebox{\plotpoint}}
\put(342.04,363.00){\usebox{\plotpoint}}
\multiput(351,363)(20.756,0.000){0}{\usebox{\plotpoint}}
\put(362.80,363.00){\usebox{\plotpoint}}
\multiput(370,363)(20.756,0.000){0}{\usebox{\plotpoint}}
\put(383.55,363.00){\usebox{\plotpoint}}
\multiput(389,363)(20.756,0.000){0}{\usebox{\plotpoint}}
\put(404.31,363.00){\usebox{\plotpoint}}
\multiput(409,363)(20.756,0.000){0}{\usebox{\plotpoint}}
\put(425.07,363.00){\usebox{\plotpoint}}
\multiput(428,363)(20.756,0.000){0}{\usebox{\plotpoint}}
\put(445.82,363.00){\usebox{\plotpoint}}
\multiput(448,363)(20.756,0.000){0}{\usebox{\plotpoint}}
\put(466.58,363.00){\usebox{\plotpoint}}
\multiput(467,363)(20.756,0.000){0}{\usebox{\plotpoint}}
\multiput(477,363)(20.756,0.000){0}{\usebox{\plotpoint}}
\put(487.33,363.00){\usebox{\plotpoint}}
\multiput(496,363)(20.756,0.000){0}{\usebox{\plotpoint}}
\put(508.09,363.00){\usebox{\plotpoint}}
\multiput(515,363)(20.756,0.000){0}{\usebox{\plotpoint}}
\put(528.84,363.00){\usebox{\plotpoint}}
\multiput(535,363)(20.756,0.000){0}{\usebox{\plotpoint}}
\put(549.60,363.00){\usebox{\plotpoint}}
\multiput(554,363)(20.756,0.000){0}{\usebox{\plotpoint}}
\put(570.35,363.00){\usebox{\plotpoint}}
\multiput(574,363)(20.756,0.000){0}{\usebox{\plotpoint}}
\put(591.11,363.00){\usebox{\plotpoint}}
\multiput(593,363)(20.756,0.000){0}{\usebox{\plotpoint}}
\put(611.87,363.00){\usebox{\plotpoint}}
\multiput(612,363)(20.756,0.000){0}{\usebox{\plotpoint}}
\multiput(622,363)(20.756,0.000){0}{\usebox{\plotpoint}}
\put(632.62,363.00){\usebox{\plotpoint}}
\multiput(641,363)(20.756,0.000){0}{\usebox{\plotpoint}}
\put(653.38,363.00){\usebox{\plotpoint}}
\multiput(661,363)(20.756,0.000){0}{\usebox{\plotpoint}}
\put(674.13,363.00){\usebox{\plotpoint}}
\multiput(680,363)(20.756,0.000){0}{\usebox{\plotpoint}}
\put(694.89,363.00){\usebox{\plotpoint}}
\multiput(700,363)(20.756,0.000){0}{\usebox{\plotpoint}}
\put(715.64,363.00){\usebox{\plotpoint}}
\multiput(719,363)(20.756,0.000){0}{\usebox{\plotpoint}}
\put(736.40,363.00){\usebox{\plotpoint}}
\multiput(738,363)(20.756,0.000){0}{\usebox{\plotpoint}}
\put(757.15,363.00){\usebox{\plotpoint}}
\multiput(758,363)(20.756,0.000){0}{\usebox{\plotpoint}}
\multiput(768,363)(20.756,0.000){0}{\usebox{\plotpoint}}
\put(777.91,363.00){\usebox{\plotpoint}}
\multiput(787,363)(20.756,0.000){0}{\usebox{\plotpoint}}
\put(798.66,363.00){\usebox{\plotpoint}}
\multiput(806,363)(20.756,0.000){0}{\usebox{\plotpoint}}
\put(819.42,363.00){\usebox{\plotpoint}}
\multiput(826,363)(20.756,0.000){0}{\usebox{\plotpoint}}
\put(840.18,363.00){\usebox{\plotpoint}}
\multiput(845,363)(20.756,0.000){0}{\usebox{\plotpoint}}
\put(860.93,363.00){\usebox{\plotpoint}}
\multiput(864,363)(20.756,0.000){0}{\usebox{\plotpoint}}
\put(881.69,363.00){\usebox{\plotpoint}}
\multiput(884,363)(20.756,0.000){0}{\usebox{\plotpoint}}
\put(902.44,363.00){\usebox{\plotpoint}}
\multiput(903,363)(20.756,0.000){0}{\usebox{\plotpoint}}
\multiput(913,363)(20.756,0.000){0}{\usebox{\plotpoint}}
\put(923.20,363.00){\usebox{\plotpoint}}
\multiput(932,363)(20.756,0.000){0}{\usebox{\plotpoint}}
\put(943.95,363.00){\usebox{\plotpoint}}
\multiput(952,363)(20.756,0.000){0}{\usebox{\plotpoint}}
\put(964.71,363.00){\usebox{\plotpoint}}
\multiput(971,363)(20.756,0.000){0}{\usebox{\plotpoint}}
\put(985.46,363.00){\usebox{\plotpoint}}
\multiput(991,363)(20.756,0.000){0}{\usebox{\plotpoint}}
\put(1006.22,363.00){\usebox{\plotpoint}}
\multiput(1010,363)(20.756,0.000){0}{\usebox{\plotpoint}}
\put(1026.98,363.00){\usebox{\plotpoint}}
\multiput(1029,363)(20.756,0.000){0}{\usebox{\plotpoint}}
\put(1047.73,363.00){\usebox{\plotpoint}}
\multiput(1049,363)(20.756,0.000){0}{\usebox{\plotpoint}}
\multiput(1058,363)(20.756,0.000){0}{\usebox{\plotpoint}}
\put(1068.49,363.00){\usebox{\plotpoint}}
\multiput(1078,363)(20.756,0.000){0}{\usebox{\plotpoint}}
\put(1089.24,363.00){\usebox{\plotpoint}}
\multiput(1097,363)(20.756,0.000){0}{\usebox{\plotpoint}}
\put(1110.00,363.00){\usebox{\plotpoint}}
\multiput(1117,363)(20.756,0.000){0}{\usebox{\plotpoint}}
\put(1130.75,363.00){\usebox{\plotpoint}}
\put(1136,363){\usebox{\plotpoint}}
\sbox{\plotpoint}{\rule[-0.600pt]{1.200pt}{1.200pt}}%
\sbox{\plotpoint}{\rule[-0.500pt]{1.000pt}{1.000pt}}%
\put(176,475){\usebox{\plotpoint}}
\put(176.00,475.00){\usebox{\plotpoint}}
\multiput(186,475)(41.511,0.000){0}{\usebox{\plotpoint}}
\multiput(195,475)(41.511,0.000){0}{\usebox{\plotpoint}}
\multiput(205,475)(41.511,0.000){0}{\usebox{\plotpoint}}
\put(217.51,475.00){\usebox{\plotpoint}}
\multiput(224,475)(41.511,0.000){0}{\usebox{\plotpoint}}
\multiput(234,475)(41.511,0.000){0}{\usebox{\plotpoint}}
\multiput(244,475)(41.511,0.000){0}{\usebox{\plotpoint}}
\put(259.02,475.00){\usebox{\plotpoint}}
\multiput(263,475)(41.511,0.000){0}{\usebox{\plotpoint}}
\multiput(273,475)(41.511,0.000){0}{\usebox{\plotpoint}}
\multiput(283,475)(41.511,0.000){0}{\usebox{\plotpoint}}
\put(300.53,475.00){\usebox{\plotpoint}}
\multiput(302,475)(41.511,0.000){0}{\usebox{\plotpoint}}
\multiput(312,475)(41.511,0.000){0}{\usebox{\plotpoint}}
\multiput(321,475)(41.511,0.000){0}{\usebox{\plotpoint}}
\multiput(331,475)(41.511,0.000){0}{\usebox{\plotpoint}}
\put(342.04,475.00){\usebox{\plotpoint}}
\multiput(351,475)(41.511,0.000){0}{\usebox{\plotpoint}}
\multiput(360,475)(41.511,0.000){0}{\usebox{\plotpoint}}
\multiput(370,475)(41.511,0.000){0}{\usebox{\plotpoint}}
\put(383.55,475.00){\usebox{\plotpoint}}
\multiput(389,475)(41.511,0.000){0}{\usebox{\plotpoint}}
\multiput(399,475)(41.511,0.000){0}{\usebox{\plotpoint}}
\multiput(409,475)(41.511,0.000){0}{\usebox{\plotpoint}}
\put(425.07,475.00){\usebox{\plotpoint}}
\multiput(428,475)(41.511,0.000){0}{\usebox{\plotpoint}}
\multiput(438,475)(41.511,0.000){0}{\usebox{\plotpoint}}
\multiput(448,475)(41.511,0.000){0}{\usebox{\plotpoint}}
\put(466.58,475.00){\usebox{\plotpoint}}
\multiput(467,475)(41.511,0.000){0}{\usebox{\plotpoint}}
\multiput(477,475)(41.511,0.000){0}{\usebox{\plotpoint}}
\multiput(486,475)(41.511,0.000){0}{\usebox{\plotpoint}}
\multiput(496,475)(41.511,0.000){0}{\usebox{\plotpoint}}
\put(508.09,475.00){\usebox{\plotpoint}}
\multiput(515,475)(41.511,0.000){0}{\usebox{\plotpoint}}
\multiput(525,475)(41.511,0.000){0}{\usebox{\plotpoint}}
\multiput(535,475)(41.511,0.000){0}{\usebox{\plotpoint}}
\put(549.60,475.00){\usebox{\plotpoint}}
\multiput(554,475)(41.511,0.000){0}{\usebox{\plotpoint}}
\multiput(564,475)(41.511,0.000){0}{\usebox{\plotpoint}}
\multiput(574,475)(41.511,0.000){0}{\usebox{\plotpoint}}
\put(591.11,475.00){\usebox{\plotpoint}}
\multiput(593,475)(41.511,0.000){0}{\usebox{\plotpoint}}
\multiput(603,475)(41.511,0.000){0}{\usebox{\plotpoint}}
\multiput(612,475)(41.511,0.000){0}{\usebox{\plotpoint}}
\multiput(622,475)(41.511,0.000){0}{\usebox{\plotpoint}}
\put(632.62,475.00){\usebox{\plotpoint}}
\multiput(641,475)(41.511,0.000){0}{\usebox{\plotpoint}}
\multiput(651,475)(41.511,0.000){0}{\usebox{\plotpoint}}
\multiput(661,475)(41.511,0.000){0}{\usebox{\plotpoint}}
\put(674.13,475.00){\usebox{\plotpoint}}
\multiput(680,475)(41.511,0.000){0}{\usebox{\plotpoint}}
\multiput(690,475)(41.511,0.000){0}{\usebox{\plotpoint}}
\multiput(700,475)(41.511,0.000){0}{\usebox{\plotpoint}}
\put(715.64,475.00){\usebox{\plotpoint}}
\multiput(719,475)(41.511,0.000){0}{\usebox{\plotpoint}}
\multiput(729,475)(41.511,0.000){0}{\usebox{\plotpoint}}
\multiput(738,475)(41.511,0.000){0}{\usebox{\plotpoint}}
\put(757.15,475.00){\usebox{\plotpoint}}
\multiput(758,475)(41.511,0.000){0}{\usebox{\plotpoint}}
\multiput(768,475)(41.511,0.000){0}{\usebox{\plotpoint}}
\multiput(777,475)(41.511,0.000){0}{\usebox{\plotpoint}}
\multiput(787,475)(41.511,0.000){0}{\usebox{\plotpoint}}
\put(798.66,475.00){\usebox{\plotpoint}}
\multiput(806,475)(41.511,0.000){0}{\usebox{\plotpoint}}
\multiput(816,475)(41.511,0.000){0}{\usebox{\plotpoint}}
\multiput(826,475)(41.511,0.000){0}{\usebox{\plotpoint}}
\put(840.18,475.00){\usebox{\plotpoint}}
\multiput(845,475)(41.511,0.000){0}{\usebox{\plotpoint}}
\multiput(855,475)(41.511,0.000){0}{\usebox{\plotpoint}}
\multiput(864,475)(41.511,0.000){0}{\usebox{\plotpoint}}
\put(881.69,475.00){\usebox{\plotpoint}}
\multiput(884,475)(41.511,0.000){0}{\usebox{\plotpoint}}
\multiput(894,475)(41.511,0.000){0}{\usebox{\plotpoint}}
\multiput(903,475)(41.511,0.000){0}{\usebox{\plotpoint}}
\multiput(913,475)(41.511,0.000){0}{\usebox{\plotpoint}}
\put(923.20,475.00){\usebox{\plotpoint}}
\multiput(932,475)(41.511,0.000){0}{\usebox{\plotpoint}}
\multiput(942,475)(41.511,0.000){0}{\usebox{\plotpoint}}
\multiput(952,475)(41.511,0.000){0}{\usebox{\plotpoint}}
\put(964.71,475.00){\usebox{\plotpoint}}
\multiput(971,475)(41.511,0.000){0}{\usebox{\plotpoint}}
\multiput(981,475)(41.511,0.000){0}{\usebox{\plotpoint}}
\multiput(991,475)(41.511,0.000){0}{\usebox{\plotpoint}}
\put(1006.22,475.00){\usebox{\plotpoint}}
\multiput(1010,475)(41.511,0.000){0}{\usebox{\plotpoint}}
\multiput(1020,475)(41.511,0.000){0}{\usebox{\plotpoint}}
\multiput(1029,475)(41.511,0.000){0}{\usebox{\plotpoint}}
\put(1047.73,475.00){\usebox{\plotpoint}}
\multiput(1049,475)(41.511,0.000){0}{\usebox{\plotpoint}}
\multiput(1058,475)(41.511,0.000){0}{\usebox{\plotpoint}}
\multiput(1068,475)(41.511,0.000){0}{\usebox{\plotpoint}}
\multiput(1078,475)(41.511,0.000){0}{\usebox{\plotpoint}}
\put(1089.24,475.00){\usebox{\plotpoint}}
\multiput(1097,475)(41.511,0.000){0}{\usebox{\plotpoint}}
\multiput(1107,475)(41.511,0.000){0}{\usebox{\plotpoint}}
\multiput(1117,475)(41.511,0.000){0}{\usebox{\plotpoint}}
\put(1130.75,475.00){\usebox{\plotpoint}}
\put(1136,475){\usebox{\plotpoint}}
\end{picture}

{\caption{Typical form of $f_{\eta'}(p)$. With increasing $\eta'$ the
abcissa shifts downwards and passes for $\eta'=1$ through the point
of inflection at $p_{\rm infl}$.}}
\end{figure}
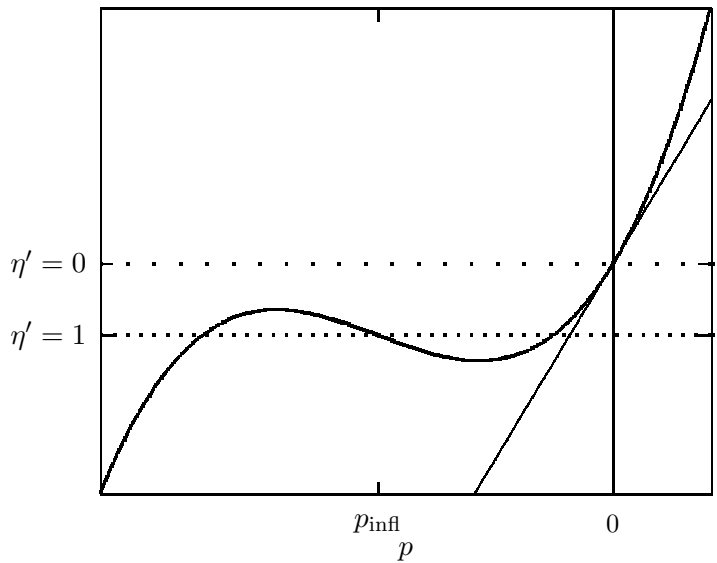

\end{document}